\documentclass[aps, prb, amsmath,amssymb,
 reprint, superscriptaddress,
]{revtex4-1}

\usepackage{graphicx}
\usepackage{dcolumn}
\usepackage{bm}
\usepackage{color}
\usepackage{longtable}
\usepackage[colorlinks=true,allcolors=blue]{hyperref}

\begin{document}
\newcommand{\micron}{$\mu$m}

\title{Nature of the bonded-to-atomic transition in liquid silica to TPa pressures}

\author{Shuai Zhang}
\email{szha@lle.rochester.edu}
\affiliation{Laboratory for Laser Energetics, University of Rochester, Rochester, New York 14623, USA}

\author{Miguel A. Morales}
\affiliation{Center for Computational Quantum Physics, Flatiron Institute, New York, New York 10010, USA}
\affiliation{Lawrence Livermore National Laboratory, Livermore, California 94550, USA}

\author{Raymond Jeanloz}
\affiliation{University of California, Departments of Earth \& Planetary Science and Astronomy, Berkeley, California 94720, USA}

\author{Marius Millot}
\affiliation{Lawrence Livermore National Laboratory, Livermore, California 94550, USA}

\author{S. X. Hu}
\affiliation{Laboratory for Laser Energetics, University of Rochester, Rochester, New York 14623, USA}

\author{Eva Zurek}
\affiliation{Department of Chemistry, State University of New York at Buffalo, Buffalo, NY 14260-3000, USA}

\date{\today}

\begin{abstract}
First-principles calculations 
and analysis of the thermodynamic, structural, and electronic properties of liquid SiO$_2$
characterize the bonded-to-atomic transition at 0.1--1.6~TPa and 10$^4$--10$^5$~K (1--7~eV), the high-energy-density regime relevant to understanding planetary interiors.
We find 
strong ionic bonds that become short-lived due to high kinetics 
during the transition, with sensitivity of the transition temperature to pressure, and our calculated Hugoniots agree with past experimental data.
These results reconcile previous experimental and theoretical findings by clarifying the nature of the bond dissociation process in early Earth and ``rocky'' (oxide) constituents of large planets.
\end{abstract}

\maketitle

\section{Introduction}

Silica (SiO$_2$), a key constituent of Earth, terrestrial (``rocky'') and even giant planets, is an important compound for theory, basic science and technology, including as a laboratory standard for high-energy-density (HED) experiments.  Its response to dynamic compression helps to determine i) how planets form through giant impacts, and ii) the high pressure–temperature material properties that control, for example, how the deep interior of planets evolve.

Starting from $\alpha$-quartz at ambient condition, SiO$_2$ goes through a series of phase transitions as pressure increases~\cite{Gillan2006}: first to coesite at 2 GPa, then to stishovite at 8 GPa, a CaCl$_2$ structure at 50 GPa, an $\alpha$-PbO$_2$ structure at 100 GPa, and a pyrite-type structure at 200 GPa.
At higher pressures (700~GPa), simulations predict a cotunnite structure (if the temperature exceeds $\sim$1000 K), or a Fe$_2$P phase, with the latter being stable to 2000 GPa (2 TPa)~\cite{Tsuchiya2011}.
In addition to the thermodynamically stable phases, a number of metastable silica polymorphs and their transformation have been studied~\cite{DUBROVINSKY2004231,Cernok2018ncomm,Shelton2018}. The dynamic response of fused silica~\cite{Tracy2018} has recently also been measured with {\it in situ} x-ray diffraction.

Developments in dynamic experiments over the past two decades have provided important constraints on the high-temperature phase diagram and properties of SiO$_2$ at 100 GPa and above. Hicks {\it et al.}~\cite{HicksPRL2006} measured temperatures and reflectivities along the Hugoniots of $\alpha$-quartz and fused silica from near the melting curve up to 1~TPa, and reported 
specific heat capacities that exceed the Dulong--Petit limit.
Kraus {\it et al.}~\cite{Kraus2012} performed shock-and-release experiments, and set criteria for vaporization of $\alpha$-quartz.
Millot {\it et al.}~\cite{MillotSci2015} conducted laser shock experiments on 
stishovite crystals
and determined the temperature-pressure-density equation of state (EOS), electronic conductivity, and melting temperature along the Hugoniot of stishovite.
McCoy {\it et al.}~\cite{McCoy2016b} used an unsteady wave method for measuring the sound velocity of fused silica shocked up to 1.1~TPa, which relies on an analytic release model for the sound velocity of the $\alpha$-quartz reference.
Li {\it et al.}~\cite{Li2018} developed a lateral release approach to continuously measure the sound velocity along the Hugoniot of $\alpha$-quartz, and calculated its Gr\"uneisen parameters to 1.45~TPa.
Recently, Guarguaglini {\it et al.}~\cite{Guarguaglini2021} designed double-shock experiments of $\alpha$-quartz and explored the EOS and two-color reflectivity of silica in the temperature-pressure regime between the Hugonoit curves of $\alpha$-quartz and stishovite.
To date, magnetically and laser driven experiments combined with first-principles molecular dynamics simulations~\cite{KD2013_aquartz,QiPoP2015,Knudson2013aerogel,McCoy2016,Marshall2019QuartzMo,Root2019fusedsilica,SjostromAIP2017,MillotSci2015} have produced a large number of Hugoniot data,
up to 6.2~TPa for $\alpha$-quartz, 0.2~TPa for silica aerogels, 1.6~TPa for fused silica, and 2.5~TPa for stishovite.
These results have established $\alpha$-quartz and fused silica as standards for impedance matching at up to 1.2~TPa.

Despite this progress, questions remain about Hugoniot temperatures~\cite{Falk2014,SjostromAIP2017} and reflectivity~\cite{QiPoP2015,ScipioniPNAS2017} estimated in laser shock experiments, as well as about changes in the structure of silica liquids~\cite{HicksPRL2006,Kraus2012,ScipioniPNAS2017,Green2018}.
Based on anomalies (i.e., minimum) in the observed heat capacity, Hicks {\it et al.}~\cite{HicksPRL2006} proposed that a temperature-induced bonded-to-atomic transition occurs near 37,000 K in liquid silica, with little variation up to pressures of about 1~TPa. 
In contrast, the heat-capacity variations were interpreted as non-dissociative changes in atomic and electronic structure in a recent computational study~\cite{ScipioniPNAS2017}.
Despite overall agreement between previous theoretical results and experimentally measured Hugoniots in the liquid regime of silica~\cite{QiPoP2015,ScipioniPNAS2017,SjostromAIP2017,Root2019fusedsilica}, the atomistic and electronic structure and their changes with temperature and pressure have not been addressed explicitly.

Understanding liquid structural changes at extreme conditions, particularly upon bonded-to-atomic transitions (e.g., molecular-to-atomic transition), not only helps to clarify phase transitions and metallization that generally occurs in materials such as hydrogen~\cite{Morales2010,Rillo2019,Celliers2018,Ohta2015,Zaghoo2016,Jiang2020H,McWilliams2016,Hinz2020PRR} 
and nitrogen~\cite{Nellis1991,Weck2017,Jiang2018,Kim2022}, 
but can also shed light on material transport properties (e.g., electrical and thermal conductivity) critical to modeling the dynamics of magma ocean and magnetic field generation in early Earth and super-Earth exoplanets~\cite{MillotSci2015,ScipioniPNAS2017,Soubiran2018,Stixrude2020}, as well as
for numerical simulations of giant impacts~\cite{Melosh2007,Stewart2020,Kraus2012,Green2018}.

The goal of this work is to provide in-depth analysis and theoretical insights about the structural changes and the nature of the bonded-to-atomic transition in liquid silica at extreme conditions by way of first-principles quantum simulations.
The manuscript is outlined as follows: 
Sec.~\ref{sec:method} provides the computational details; Sec.~\ref{sec:result} shows our results that elucidate the transition from various perspectives; and Sec.~\ref{sec:conclusions} 
discusses questions of interest for future studies. 

\section{Methods}\label{sec:method}
We conducted molecular dynamics (MD) simulations of silica along five different isochores based on Kohn--Sham density functional theory (DFT)~\cite{ks1965}.
The density and temperature ranges that we have considered are between 2.65--7.95~g/cm$^3$ (1--3 times the ambient density of $\alpha$-quartz) and 5000--100,000~K. The corresponding temperature and pressure conditions are around those experimentally probed along the Hugoniots of $\alpha$-quartz and fused silica~\cite{HicksPRL2006}.
The simulation cells contained 8 or 24 formula units (f.u.) of SiO$_2$, except for certain cases (indicated with yellow pentagons in Fig.~\ref{fig:hugtrhop}(a)) where we used 64-f.u. (192-atom) cells.

By using a Nos\'{e} thermostat~\cite{Nose1984}, we generated a canonical (constant-$NVT$, where $N$, $V$, and $T$ are respectively the number of atoms, volume, and the equilibrium temperature of the system) ensemble at each temperature-density condition of interest that typically consisted of a DFT-MD trajectory of at least 2000 steps for the 8--24~f.u. and 5000--25,000 steps for the 64-f.u. simulations (timestep is 0.2--0.5~fs).
When analyzing the EOS, we threw away the beginning part (20\%) of each MD trajectory to ensure the reported EOS represents that under thermodynamic equilibrium.
Ion kinetic contributions to the EOS are manually included by following an ideal gas formula (i.e., internal energy $E_\textrm{ion kin.}=3Nk_\textrm{B}T/2$ and pressure $P_\textrm{ion kin.}=3Nk_\textrm{B}T/V$, where $k_\text{B}$ is the Boltzmann constant), while all other contributions (ion-ion, ion-electron, and electron-electron interactions and the electron kinetic term) are calculated explicitly in the
Vienna \textit{Ab initio} Simulation Package ({\footnotesize VASP})~\cite{kresse96b}.
Electrons are enforced to follow a Fermi-Dirac distribution with the temperature equal to that of the ions~\cite{mermin1965}. 

DFT calculations were done by using the projector augmented wave (PAW)~\cite{Blochl1994} method, plane-wave basis sets, and exchange-correlation (XC) functionals under the local density approximation (LDA)~\cite{Ceperley1980,Perdew81}, which are implemented in {\footnotesize VASP}.
We use the hardest available PAW potentials with a core radius of 1.6 Bohr and four electrons treated as the valence for silicon, and a 1.1-Bohr core and six valence electrons for oxygen. We used the $\Gamma$ point to sample the Brillouin zone during the calculations.

We cross checked the calculations by using Perdew-Burke-Ernzerhof (PBE)-type or GW-type pseudopotentials and XC functionals based on the generalized gradient approximation (GGA, such as the PBE~\cite{Perdew96} or  Armiento-Mattsson (AM05)~\cite{AM05a} types) at selected conditions, in order to determine the methodological error of our results. We use a denser 4$\times$4$\times$4 $k$-point mesh to check the finite size effects on our results.

\section{Results}\label{sec:result}
\subsection{Equation of state, shock Hugoniot, and thermodynamic properties}

\begin{figure}[ht]
\centering
\includegraphics[width=0.9\linewidth]{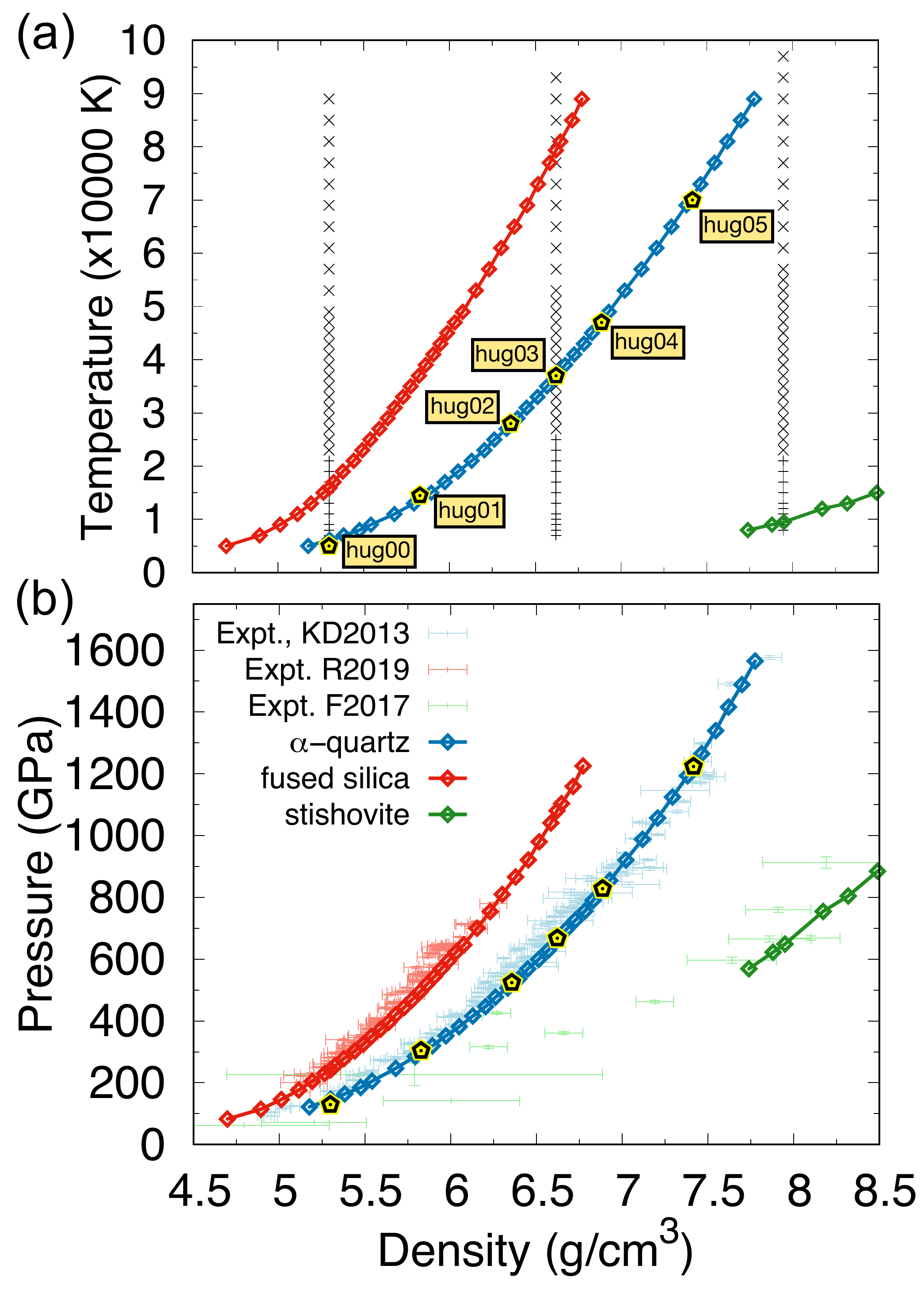}
\caption{(a) Temperature-density and (b) pressure-density plots of the shock Hugoniot of SiO$_2$ in the initial form of fused silica (red), $\alpha$-quartz (blue) and stishovite (green). 
In (a), black symbols denote the conditions of the first-principles simulations for the EOS (`+' and `x' represent 24-f.u. and 8-f.u. simulations, respectively). In (b), experimental Hugoniots from Refs.~\onlinecite{KD2013_aquartz,Root2019fusedsilica,Furnish2017_stishovite} are shown for comparison. 
Yellow pentagons denote the near-Hugoniot conditions (``hug00--05'') at which 192-atom simulations were performed.}
\label{fig:hugtrhop}
\end{figure}

Based on conservation of mass, momentum, and energy across the shock front, the state of a material under steady shock is generally related to its initial state through the Rankine--Hugoniot equation
$E-E_i+(P+P_i)(V-V_i)/2=0$,
where $(E, P, V)$ denote the internal energy, pressure, and volume of the material in the shocked state and $(E_i, P_i, V_i)$ are the corresponding values at the initial unshocked state.
This equation defines the locus of states that the material can reach when being shocked, which is known as the Hugoniot.

Numerically, one way to determine the Hugoniot is by starting from the EOS calculations on a temperature-density grid. 
In this work, 
the initial states are estimated by starting from DFT calculations at the desired densities (2.20, 2.65, and 4.29~g/cm$^3$ for fused silica, $\alpha$-quartz, and stishovite, respectively)~\footnote{We perform ground-state DFT calculations under LDA for $\alpha$-quartz at the fixed density of 2.65~g/cm$^3$ and use the resultant values in internal energy (-26.089~eV/SiO$_2$) and pressure (-0.825 GPa) for $E_i$ and $P_i$ when calculating the Hugoniot of $\alpha$-quartz. We use the same $E_i$ value and $P_i=0$ to respectively approximate the initial internal energy and pressure of fused silica at 2.2~g/cm$^3$. This is a reasonable guess because $\alpha$-quartz and fused silica are common polymorphs of SiO$_2$ at ambient condition ($P\approx0$), indicating the minimum of their respective $E(V)$ cold curves are similar to each other (so that their common tangent, if exists, has zero slope). We have also tried a slightly higher value (by 20 meV/SiO$_2$) for $E_i$ of fused silica to approximate possible differences from other sources (e.g., vibration and nuclear quantum effects), and the resultant Hugoniots remain similar. The good agreements with experimental Hugoniots (Fig.~\ref{fig:hugtrhop}(b)) also suggest that our initial conditions are reasonable for estimating the Hugoniots of liquid silica. For stishovite with the initial density of 4.29~g/cm$^3$, we use the same method to estimate its initial energy and note that we get a similar value (-26.079~eV/SiO$_2$) from zero-pressure DFT calculations, using which as $E_i$ the stishovite Hugoniot remains the same (the difference is less than 0.2\%).},
then we consider each isotherm with temperature $T$ and fit the pressure and energy data along the isotherm as functions of density by using cubic splines~\footnote{We use at least three data points in each fitting.}, and the density $\rho$ at which the energy term $[E-E_i]$ equals the pressure term $[(P+P_i)(V_i-V)/2]$ defines the Hugoniot, which has definitive values in $T, \rho, P$ and $E$. We also calculate the shock velocity $u_s$ and particle velocity $u_p$, relevant to shock experiments, by $u_s^2=\xi/\eta$ and $u_p^2=\xi\eta$, where $\xi=(P-P_i)/\rho_i$ and $\eta=1-\rho_i/\rho$. 
This approach has been used previously for calculating the Hugoniot of several other materials~\cite{Zhang2017ch,Zhang2018ch,Zhang2018b,Zhang2019bn,Zhang2020b4c1,Zhang2020b4c2,Millot2020},
and was found to produce consistent Hugoniots with other computational methods, such as 
progressive determination by running a large number of EOS calculations around the Hugoniot curve~\cite{Shamp2017,lepape2013}

In order to cross check the validity of the Hugoniot results based on the relatively sparse temperature-density grid, we have recalculated the Hugoniot of $\alpha$-quartz by performing 2D interpolation of the pressure and energy data as functions of $(T,\rho)$ and then determined the conditions at which the function $\mathcal{H}(\rho,T)=E-E_i+(P+P_i)(V-V_i)/2$ equals zero. We have also made tests by using a partial set of our EOS data (by excluding the 6.62~g/cm$^3$ isochore). The Hugoniots obtained from the different methods differ by no more than 3.5\%. Such small differences in the Hugoniot do not affect our comparisons with experiments or other discussions in the following sections.
{\color{black}We have performed additional EOS calculations to 41,000~K along the 9.27-g/cm$^3$ isochore by using similar settings as the lower-density ones. Adding these data to our EOS table does not affect the calculated Hugoniots but extends our stishovite Hugoniot from 9500 to 27,400~K.}

Figure~\ref{fig:hugtrhop} shows our shock Hugoniots on an EOS grid at 4.5--8.5~g/cm$^3$. 
The overall agreement with experimentally measured $P$--$\rho$ Hugoniots for fused silica, $\alpha$-quartz, and stishovite~\cite{KD2013_aquartz,Root2019fusedsilica,Furnish2017_stishovite} 
suggests our calculations and results on liquid silica are reliable.

\begin{figure}[ht]
\centering
\includegraphics[width=0.9\linewidth]{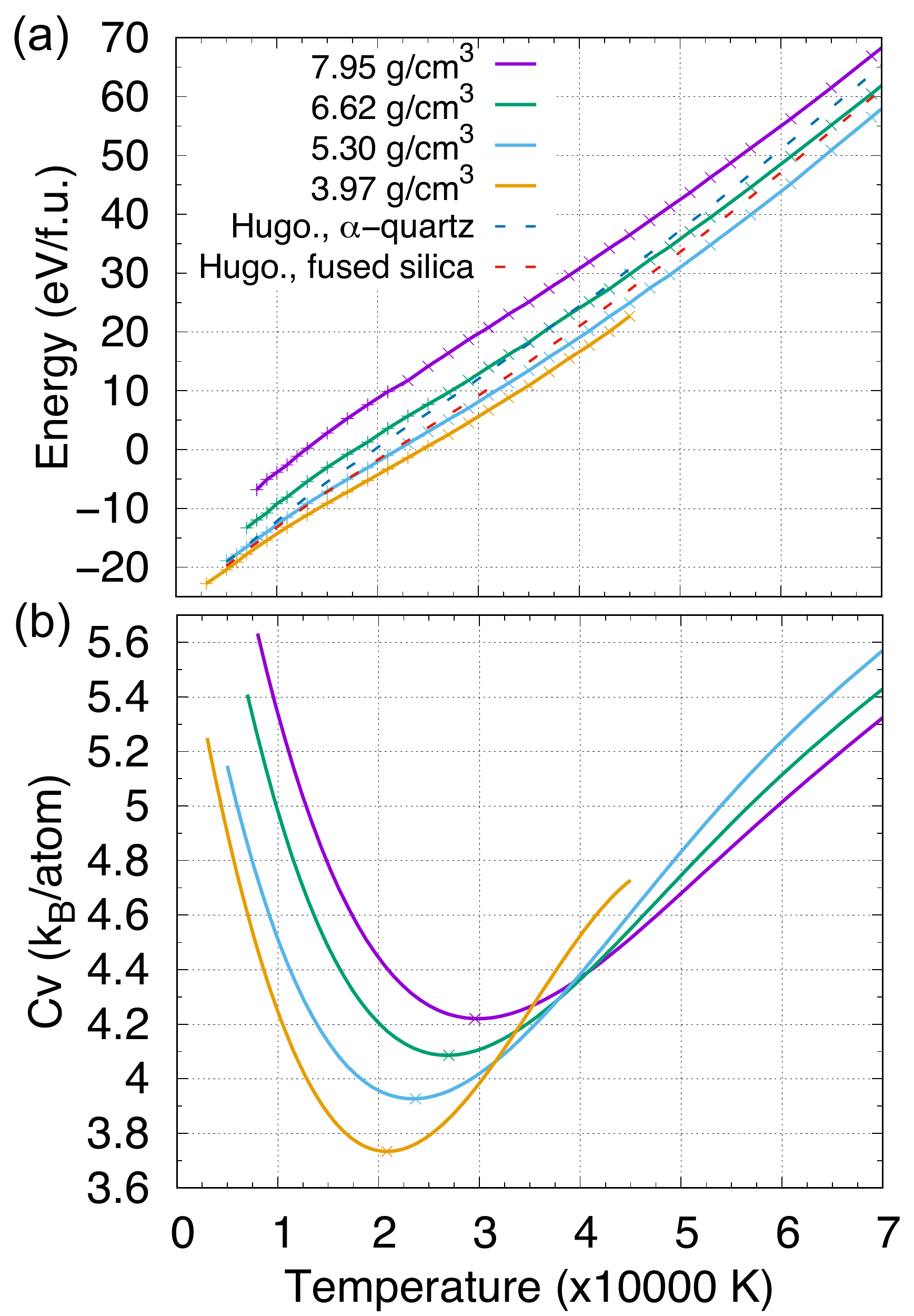}
\caption{(a) Internal energy $E(T)$ and (b) specific heat capacity $C_V(T)$ along four different isochores. In (a), `+' and `x' symbols denote results from 24-f.u. and 8-f.u. simulations, respectively; $E(T)$ profiles along two different Hugoniots are shown for comparison. In (b), the minimum of each curve (used to define the condition of anomaly in heat capacity) is shown with a cross.}
\label{fig:ecv}
\end{figure}

\begin{figure}[ht]
\centering
\includegraphics[width=0.9\linewidth]{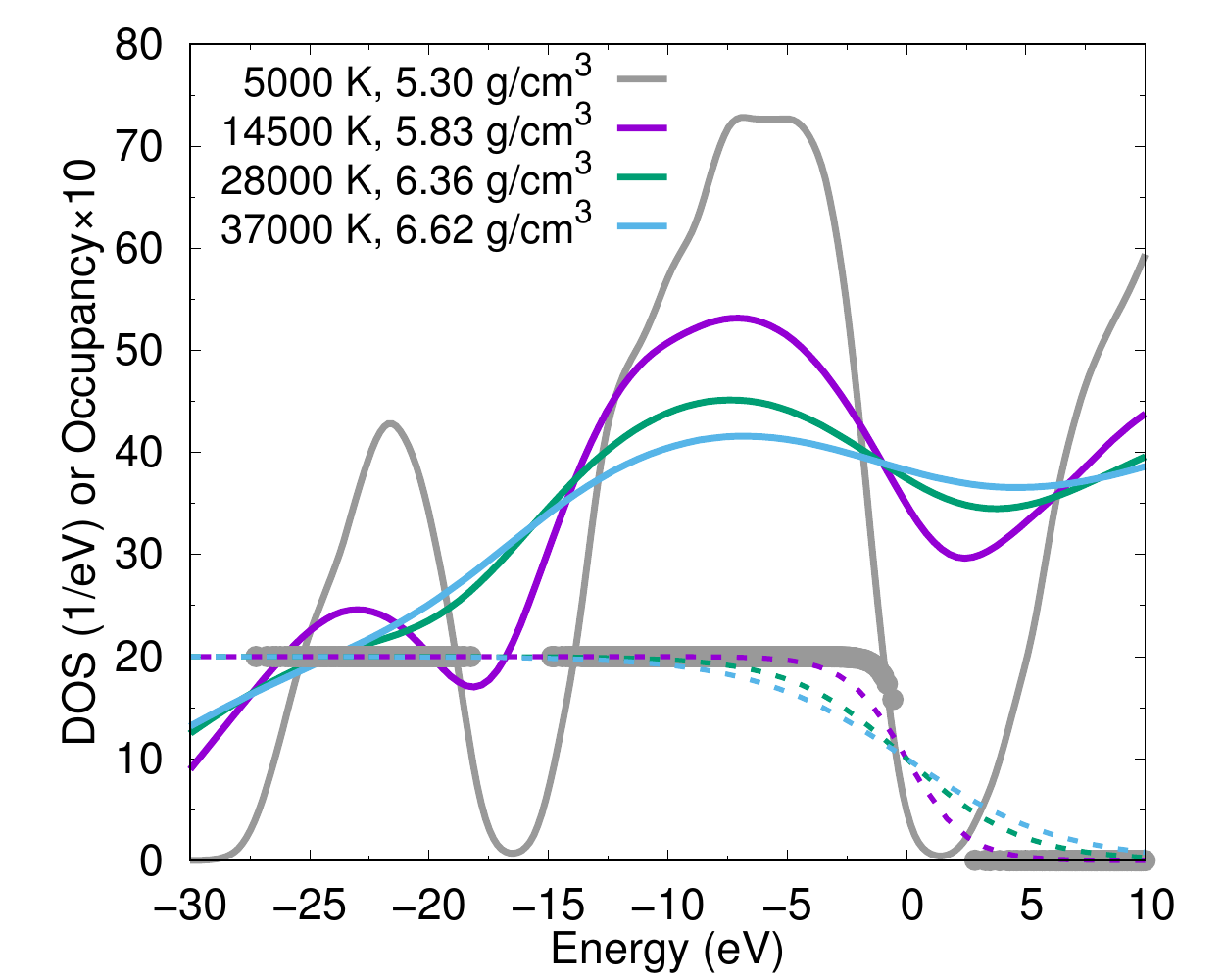}
\caption{Density of states (DOS, solid curves) and Fermi occupancy (multiplied by a factor of 10, dashed curves) of electronic states from 192-atom-cell simulations of liquid silica at different near-Hugoniot conditions (hug01--03 as labeled in Fig.~\ref{fig:hugtrhop}).  Results from a solidified structure (hug00 in Fig.~\ref{fig:hugtrhop}) are shown (in grey curve and symbols) for comparison. All profiles are aligned at 0 eV (the highest occupied state for ``hug00'' or the Fermi level for other cases). A 2.5-eV bandgap exists in the solidified structure (evidently shown by the discontinuity in Fermi occupancy) and is gradually filled in the liquid states with increasing temperature. A smearing technique (with the broadening parameter equal to the corresponding temperature) was employed to ensure smoothness of the DOS results.}
\label{fig:dos}
\end{figure}

In order to elucidate the bonded-to-atomic transition, we have firstly looked into the thermodynamic properties by plotting the internal energies and the heat capacities as functions of temperature.
Figure~\ref{fig:ecv} shows the results along four different isochores (solid curves) and along the Hugoniots (dashed curves).
The energy smoothly increases with temperature, while the heat capacity curve has an anomaly (i.e., local minimum), around which the values of $C_V$ are larger.
These observations imply a likely higher-than-first-order transition.
$C_V$ anomalies were similarly reported in previous experimental~\cite{HicksPRL2006} and theoretical~\cite{ScipioniPNAS2017,Green2018} studies of liquid silica.
Such high values of $C_V$ were also reported for other materials (e.g., hexagonal close-packed iron at high temperatures and pressures~\cite{Alfe2001} and shock melted magnesium oxide~\cite{McWilliamsScience2012} or diamond~\cite{EggertNphys2010}) and explained~\cite{Kraus2012,ScipioniPNAS2017} by the large degrees of freedom in the melt relative to the solid~\footnote{Note that the excess heat capacity was mistakenly interpreted in~\cite{Green2018} as a result of anharmonic vibration by referencing to a previous study on iron~\cite{Alfe2001}. Actually, Alf\`e {\it et al}.~\cite{Alfe2001} clearly stated that this is due to the electronic thermal excitation and that the anharmonic contribution to heat capacity is small.}.
{\color{black}The anomaly in $C_V$ is a result of joint ion and electron thermal effects.
First, the emergence of large $C_V$ (higher than $3k_\text{B}/\text{atom}$, the ``Dulong--Petit'' limit) upon melting is because of the combined vibration, rotation, and translational motion of local atomic pairs and clusters that form due to the loss of crystal symmetry.
With increasing temperature, bonds break frequently as kinetic motion increases, which effectively costs less energy to compress them.
Therefore, the ion thermal contribution to $C_V$ decreases, and eventually approaches the ideal-gas value of $1.5k_\text{B}/\text{ion}$ in the limit of infinitely high temperatures~\footnote{This trend is in accord with various theories for ion thermal free energies such as the Cowan model~\cite{leos1qeos,Benedict_2014,Zhang2018ch}.}; meanwhile, the electron contribution to $C_V$ increases with temperature due to thermal excitation, which exceeds the ion thermal part and induces a turnover in the heat capacity profile (shown with crosses in Fig.~\ref{fig:ecv}(b)).

The increasingly significant role of the electron thermal effects is closely related to the conductive nature of liquid silica, as shown in Fig.~\ref{fig:dos}.
While a bandgap exists in SiO$_2$ solids (shown by the discontinuity in Fermi-occupancy between 0--2.5~eV with grey symbols in Fig.~\ref{fig:dos}), it closes and a pseudogap forms at the Fermi level at higher temperatures when the system liquifies (purple curve in Fig.~\ref{fig:dos}, previously also found in MgO and MgSiO$_3$ in a computational study~\cite{Soubiran2018}), which is eventually filled at 28,000~K or above (green and blue curves).
We note that a similar behavior was found in hydrocarbons~\cite{Zhang2018ch} and is believed to be associated with metallization of the system, and also in boron carbide (B$_4$C) as a result of the disappearance of mid-range order and molecular motifs within the liquid~\cite{Shamp2017}.}

According to our DFT-MD data, the anomaly in $C_V(T)$ occurs at 2--3$\times10^4$~K along the isochores of 4--8~g/cm$^3$.
Under the same criteria as that used by Hicks {\it el al.}~\cite{HicksPRL2006}, these are the temperatures of chemical bond dissociation. 
Our data suggest the bonded-to-atomic transition occurs at lower temperatures and is more sensitive to pressure than the previous estimates based on laser-driven Hugoniot measurements (black line-diamond curve in Fig.~\ref{fig:phase}).
In order to understand these differences, a closer examination 
of the physics in atomistic and electronic levels is required. We choose five different conditions nearly along the $\alpha$-quartz Hugoniot and perform calculations with much larger simulation cells and more in depth analysis of the structural, electronic, and thermodynamic properties. The results are presented in the following sections.

\subsection{Structural evolution at Hugoniot conditions}
\begin{figure}[ht]
\centering
\includegraphics[width=0.9\linewidth]{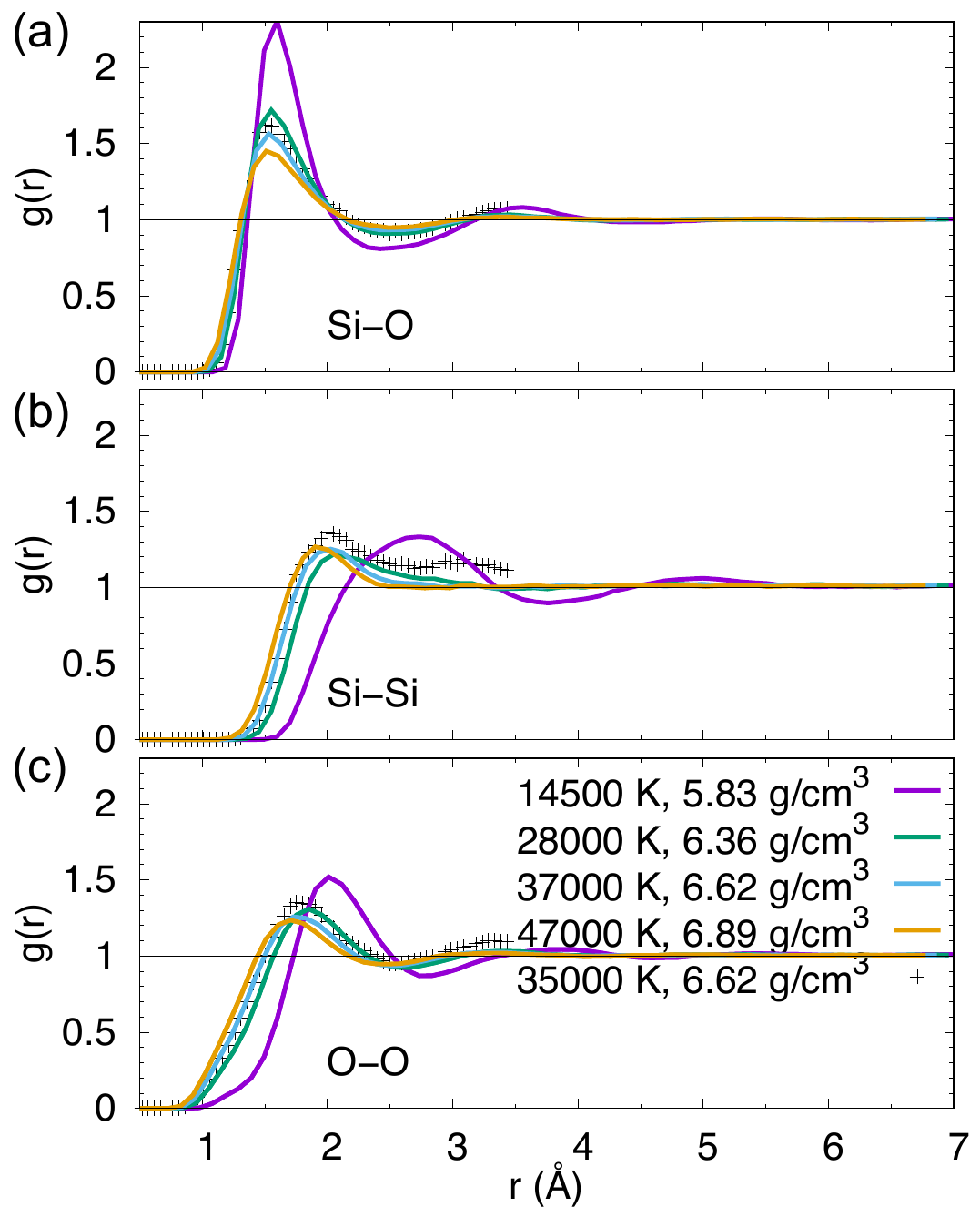}
\caption{Pair correlation functions of liquid silica at different conditions near the Hugoniot of $\alpha$-quartz: four 192-atom simulations (colored curves, corresponding to hug01-04 as labeled in Fig.~\ref{fig:hugtrhop}) and a 24-atom simulation (`+').
}
\label{fig:gr}
\end{figure}

\begin{figure*}[ht]
\centering
\includegraphics[width=0.8\linewidth]{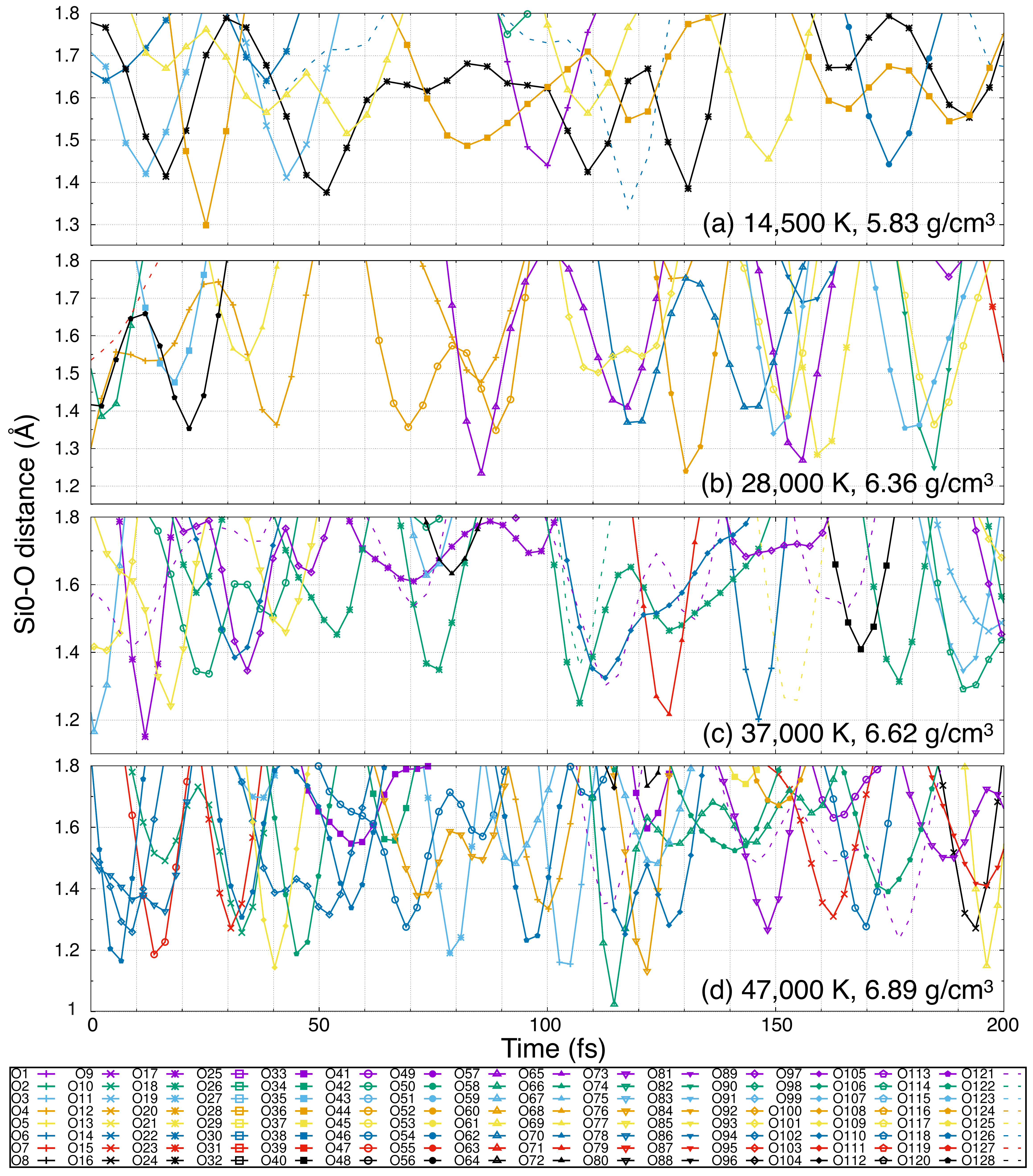}
\caption{Interatomic distances, between a Si atom (``Si0'') and all O atoms (``O1--O128'') in a 192-atom cell, as a function of time (in windows of 200 fs) during simulations at four near-Hugoniot conditions (hug01-04 as labeled in Fig.~\ref{fig:hugtrhop}).}
\label{fig:bond}
\end{figure*}

\begin{figure}[ht]
\centering
\includegraphics[width=1.0\linewidth]{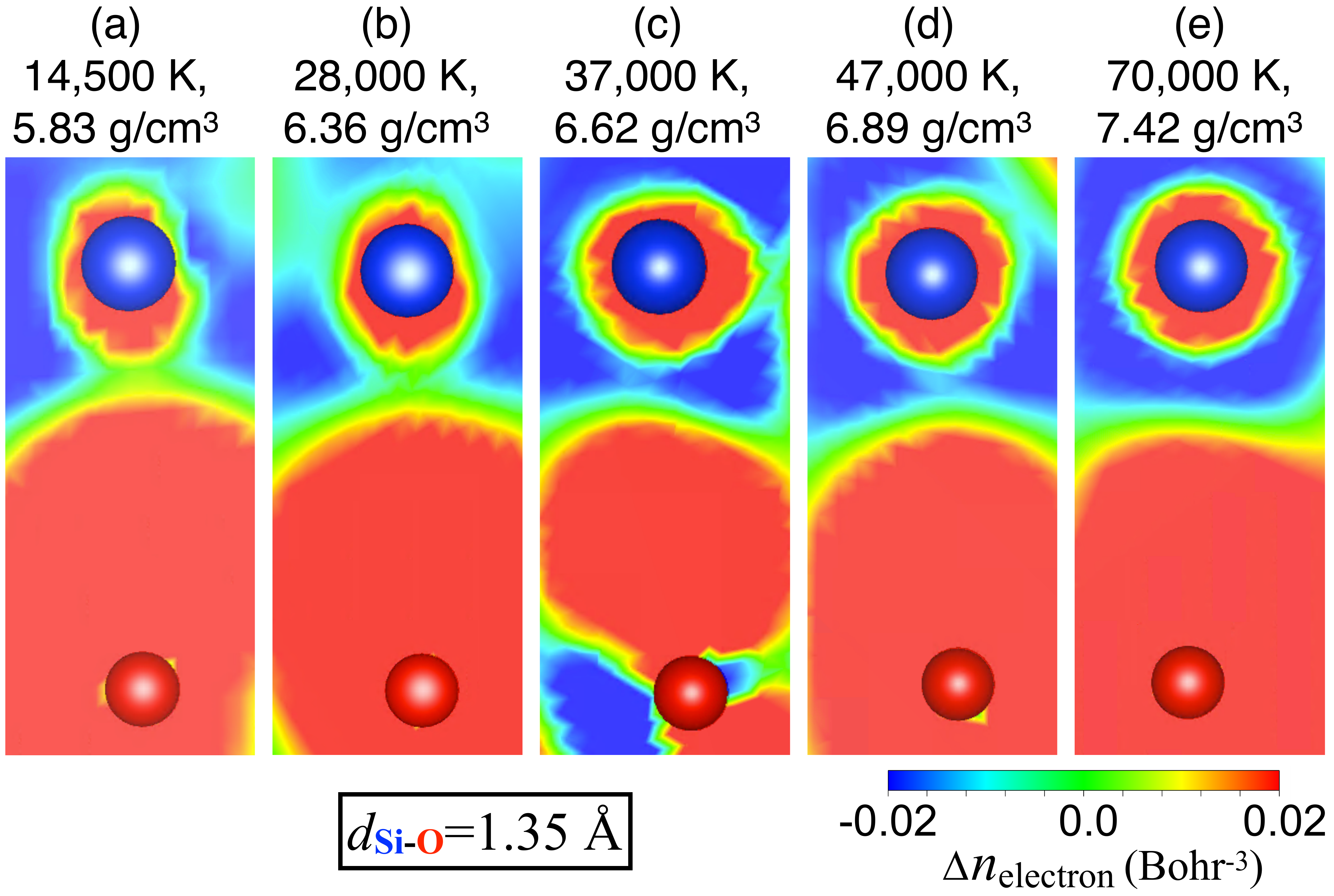}
\caption{Electron density distributions in planes around Si-O pairs from 192-atom simulations at five near-Hugoniot conditions (hug01--05 as labeled in Fig.~\ref{fig:hugtrhop}). The blue and red spheres denote the Si and O atoms, respectively, whose distances are 1.35~\AA~in all cases. The colormap displays the difference between the calculated valence electron density and the superposition of proto-atomic values.}
\label{fig:rhoe}
\end{figure}

An often useful way to describe the structure of a condensed fluid 
is by looking at the pair correlation function
$g(r)$, which is defined by the ratio between the time-averaged number density of atoms at distances $r$ from a given atom and that in an ideal gas of the same density (i.e., the average number density of the system)~\cite{AllenTildesley1987}.

\begin{figure}[ht]
\centering
\includegraphics[width=0.9\linewidth]{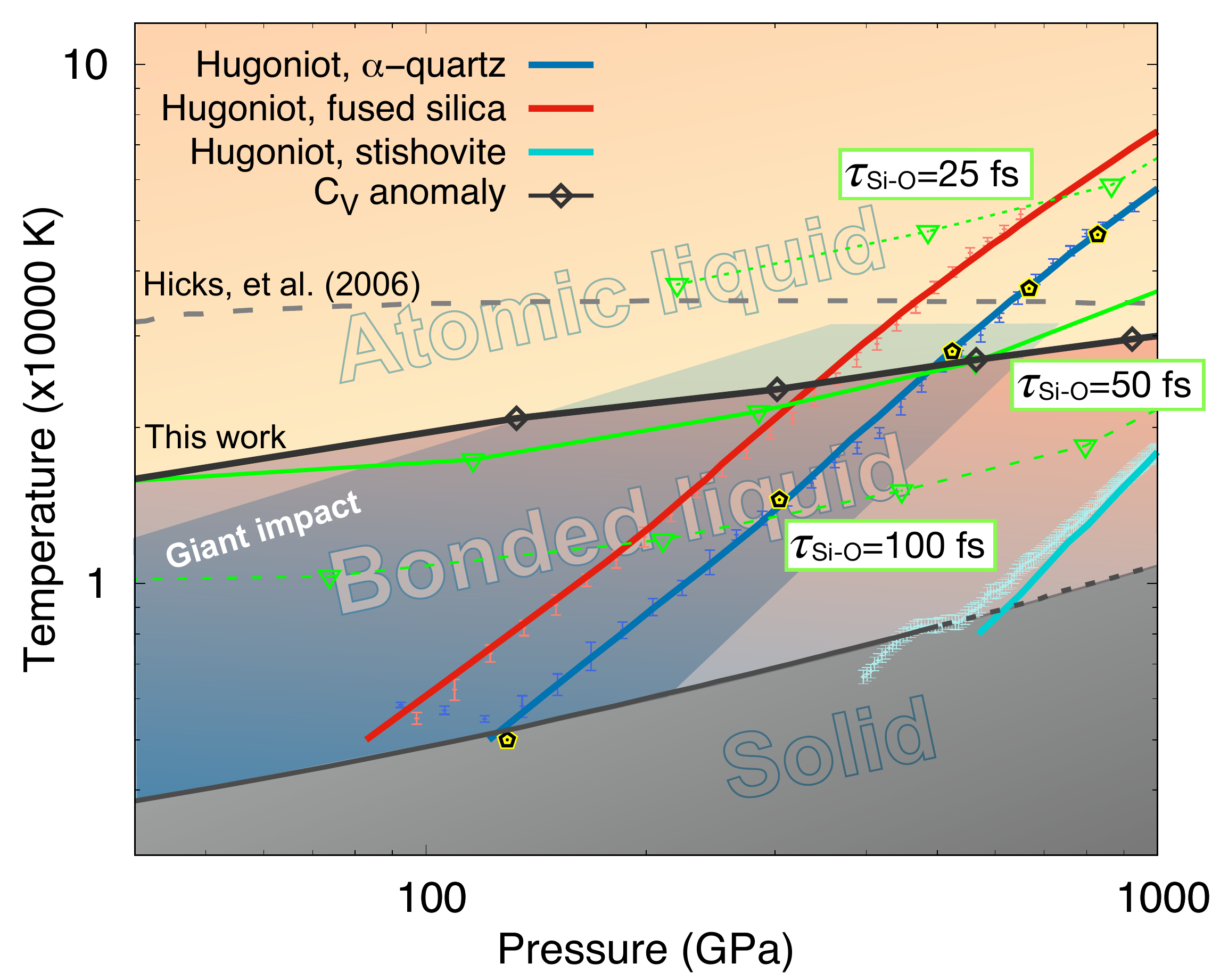}
\caption{Phase diagram of SiO$_2$ featuring the bonded-to-atomic liquid transition determined in this work (black curve with diamond symbols) as compared to a previous estimation (Hicks {\it et al.}\cite{HicksPRL2006}, grey dashed curve). Also shown are the conditions for three values of the Si--O bond lifetime (green line-triangles), Hugoniots of silica from this work (darker-colored curves in red, blue, and turquoise for fused silica, $\alpha$-quartz, and stishovite, respectively)
in comparison to experiments~\cite{HicksPRL2006,MillotSci2015} (lighter-colored symbols),
the melting curve (solid: measured; dashed: extrapolated) from Millot {\it et al.}~\cite{MillotSci2015}, and the conditions of interest (blue shaded area) to giant impacts~\cite{CanupIracus2004}. 
Yellow pentagons correspond to the near-Hugoniot conditions ``hug00--04'' labeled in Fig.~\ref{fig:hugtrhop}.
}
\label{fig:phase}
\end{figure}

Figure~\ref{fig:gr} shows the $g(r)$ results for Si--O, Si--Si, and O--O from 192-atom simulations at several different conditions along the $\alpha$-quartz Hugoniot, in comparison with that from simulations using a smaller 24-atom cell.
All $g(r)$ results show peak-valley features with tails approaching unity, typical of that in fluids,
except for the lowest temperature and density condition ``hug00'' (indicated in Fig.~\ref{fig:hugtrhop}) where we observe more structure
that originates from crystallization of the simulated structure~\footnote{{\color{black}At 5000~K and 5.30~g/cm$^3$, we found the system stabilizes into a structure that is dominated by chains of edge-shared octahedrons, with each nearby pair of SiO$_6$ units from neighbored chains sharing an O atom.}}
near the solid-liquid phase boundary (Fig.~\ref{fig:phase}). 
With increasing temperature and density, the primary peak in $g(r)$ drops in height and sharpness and shifts closer to $r=0$, as a result of thermal broadening in spatial distributions and increased compression of the system. 
The features are fully captured in the 192-atom but not by the 24-atom simulations, indicating the importance of using large cells for detailed structural analysis.
Although the small cells are sufficient in producing converged EOS data at high temperatures, much larger ones are required to understand the structure of the fluids.

Moreover, our results show clear differences between the $g(r)$ profile at 14,500 K (``hug01'') and higher-temperature ones (``hug02--05'') in appearance and values at $r<5$~\AA, 
while variations among the higher-temperature profiles between 28,000 and 70,000 K are small.
This suggests the microscopic structure of liquid silica changes qualitatively over the range of 14,500--28,000 K.
This temperature (approximately 1--3 eV) is comparable to that of typical chemical bond energies (1.5--11.1 eV)~\cite{chembondenergy},
which implies the chemical bonds in liquid silica could be subject to breaking due to the large kinetic energy.

We have therefore calculated the interatomic distances $d_\text{Si-O}$ between a randomly selected Si atom and all O atoms in the 192-atom simulation cell and monitored their changes with time. The results at four different conditions (``hug01--04'') and in a window of 200~fs
are shown in Fig.~\ref{fig:bond}(a)--(d). At 14,500 K, only a few (less than 10) oxygen atoms enter the window and some of them stay for long duration, whereas at $T\ge28,000$~K a much larger (by more than 2$\times$) number of oxygen atoms come in and out of the window, more so at higher temperatures.
This suggests that chemical bonds between Si and O both break and form  more readily (or have a shorter lifetime) in liquid silica at higher temperatures and pressures.

\begin{figure}[ht]
\centering
\includegraphics[width=1.0\linewidth]{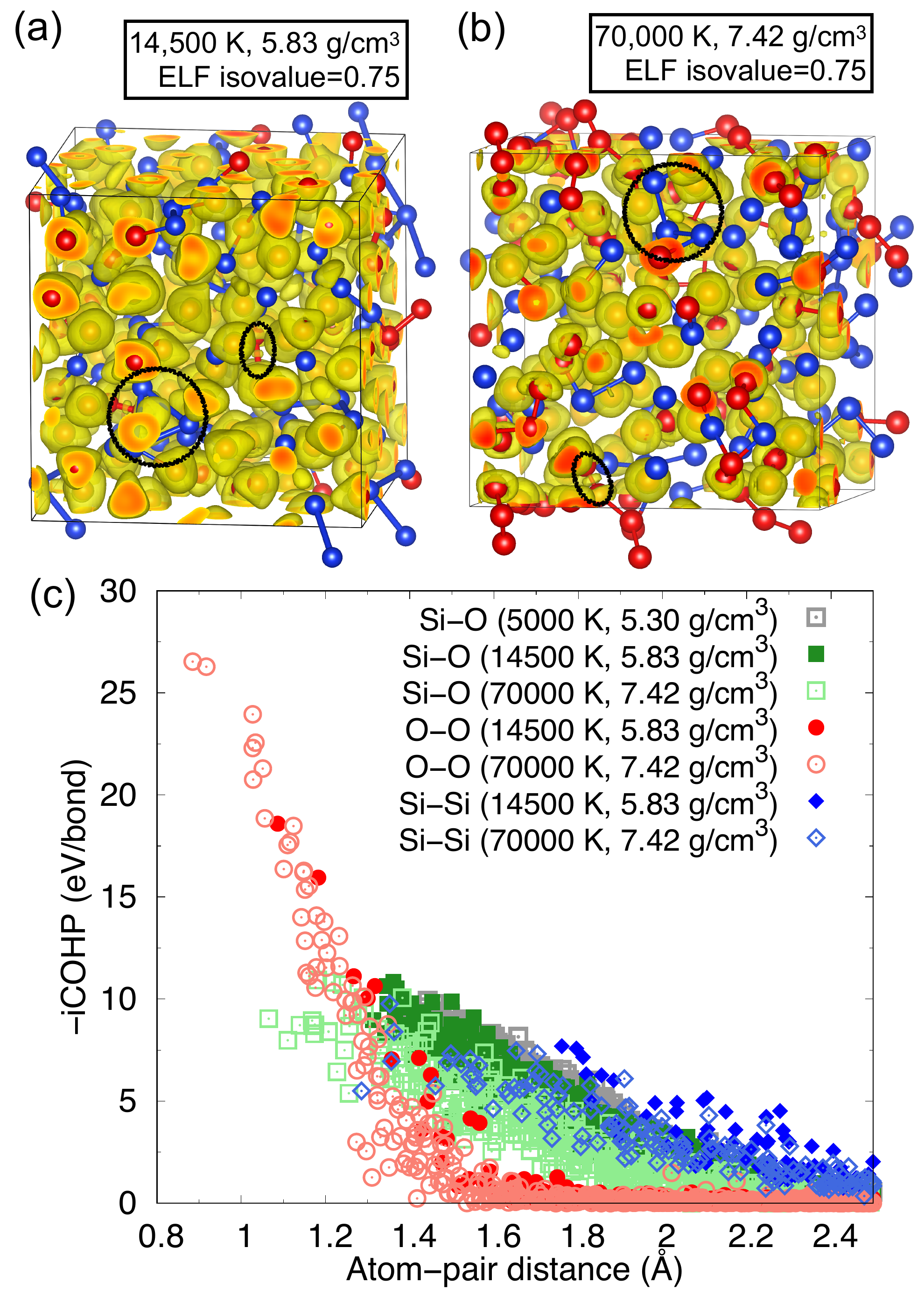}
\caption{(a--b) Electron localization function and (c) the negative of the crystal orbital Hamilton population integrated to the Fermi level (-iCOHP, which characterizes the bond strength) for atomic pairs as functions of inter-atomic distances for structure snapshots from 192-atom simulations at different near-Hugoniot conditions (hug01 and 05 as labeled in Fig.~\ref{fig:hugtrhop}). In (a) and (b), black circles denote examples of covalent bonds between homo-species (a O--O pair and a Si--Si--Si cluster). In (c), the results from a solidified structure (hug00 in Fig.~\ref{fig:hugtrhop}) that is dominated by ionic Si--O bonds are denoted by grey symbols for comparison.}
\label{fig:bonding}
\end{figure}

\begin{figure}[ht]
\centering
\includegraphics[width=1.0\linewidth]{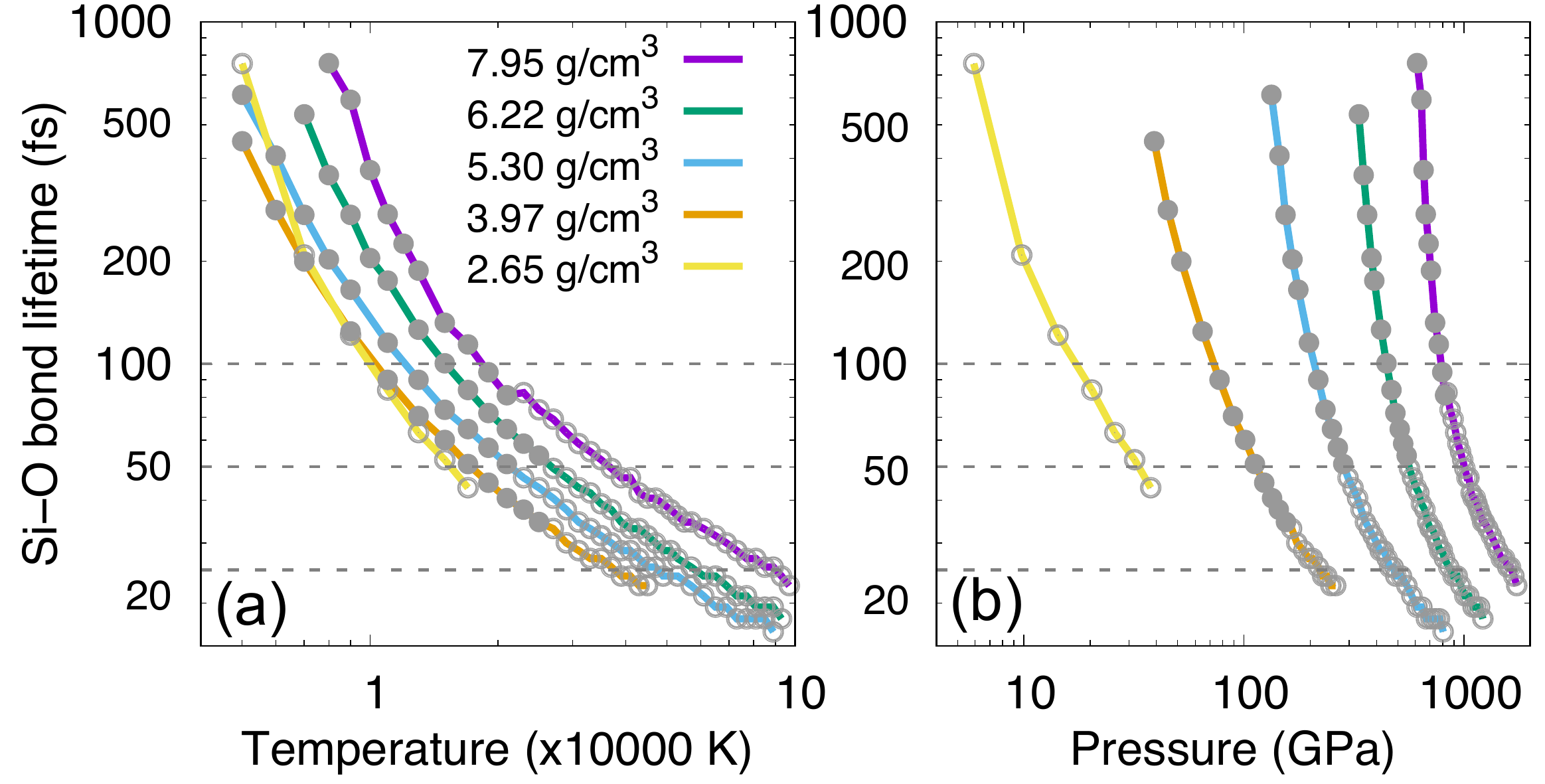}
\caption{Lifetime of Si-O bonds as a function of (a) temperature or (b) pressure along five different isochores. Open and filled circles denote simulations using 8-f.u. and 24-f.u. cells, respectively. Lifetime values of 100, 50, and 25 fs are shown with dashed horizontal lines.}
\label{fig:bondlifetime}
\end{figure}

In order to further clarify these points, we have taken a near-equilibrium snapshot of each of the simulations at different near-Hugoniot conditions ``hug01--05'', performed a self-consistent field calculation of the valence electron density at the corresponding temperature, and compared with the proto-atomic values~\footnote{The proto-atomic values of the electron density are reconstructed within the PAW method in the {\footnotesize VASP} code.
}. The resultant colormaps of the electron density difference $\Delta n_\text{electron}=n^\text{SCF}-n^\text{proto}$ at the five conditions are shown in Fig.~\ref{fig:rhoe}(a)--(e), highlighting a planar region surrounding a local Si-O structural unit in each scenario.
{\color{black}The results show that the electron density in the region between Si and O is slightly higher or similar to that of proto-atomic values at 14,500--28,000 K (yellowish green--green colors in (a)--(b)), and then becomes increasingly depleted at higher temperatures (more blueish from (c) to (e)). 
The absence of a significant gain in electron density between Si and O is indicative of the ionic nature of the bonds, and the increasing depletion of density between the atoms at higher temperatures is suggestive of a facile transition of the system from a bonded to an atomic fluid.

The electron localization function (ELF)~\cite{ELF1994Natur.371..683S} identifies regions of space that can be associated with electron pairs. Therefore, high values of the ELF are characteristic of covalent bonds or lone pairs. 
Figure~\ref{fig:bonding}(a)--(b) show ELF results for selected snapshots (same as the corresponding ones shown in Fig.~\ref{fig:rhoe} for electron densities) at two different conditions (hug01 and hug05 as labeled in Fig.~\ref{fig:hugtrhop}).
The plot shows large ELF values around oxygen, but not around silicon or between Si--O. A few regions possess large ELF values between pairs of oxygen atoms, or silicon clusters. This suggests the Si--O bonding in silica is ionic, whereas some covalent homonuclear bonds are formed at the conditions studied here.
The strength of the bonds, defined by the -iCOHP (the negative of the crystal orbital Hamilton population integrated to the Fermi level)~\cite{pCOHP.Deringer2011,lobster2016,lobster2020}, follows the same trends 
at different temperature-pressure conditions and ranges up to 12~eV/bond and 27~eV/bond for Si--O and O--O, respectively, depending on the interatomic distance (see Fig.~\ref{fig:bonding}(c)).
In comparison, the covalent bonds between silicon atoms are weaker (up to 8--10~eV/bond), particularly at lower temperatures and pressures, but they are still stronger than typical Si--Si bonds at ambient conditions (approximately 3.5~eV~\cite{chembondenergy} for single bonds with an average length of 2.34~\AA~\cite{singleSibond}) as chances are higher that atoms come closer to each other at higher temperatures and densities.
Figure~\ref{fig:bonding}(c) also shows that the bond strengths decrease with temperature. 
This is consistent with our calculated Mulliken and L\"owdin charges~\cite{Ertural2019}, which decreases from approximately $+2$ for Si and $-1$ for O in the solidified structrue (``hug00'') to around $+1$ for Si and $-0.5$ for O in the liquid states at higher temperatures and densities (``hug01--05'').
The decrease in charge therefore weakens the bonds formed by electrostatic forces, lowers the bond energy, and also contributes to the decrease in $C_V$ before it is taken control by electron thermal effects. }

We have also calculated the lifetime of Si-O bonds from the DFT-MD simulations in order to understand the kinetic effects.
The calculation is done by defining a function $F(t, r_\text{cutoff})$, which represents the probability that Si--O bonds (defined by Si--O pairs that satisfy $d_\text{Si-O}<r_\text{cutoff}$, where $r_\text{cutoff}$ is set to 2.4~\AA, the approximate position of the first valley in the pair correlation function for Si--O shown in Fig.~\ref{fig:gr}(a)) persist at time $t$.
The probability function is generally observed to be exponentially decaying with time by following $F(t, r_\text{cutoff})=\exp(-t/\tau)$, where $\tau$ is the bond lifetime. We can therefore calculate the value of $\tau$ by  fitting the probability function to the simulation time in each of the temperature-density conditions at which we have performed DFT-MD calculations.

Figure~\ref{fig:bondlifetime} shows our results for the Si--O bond lifetime along five different isochores~\footnote{Here, since we only need to count the nearest Si--O pairs ($d$ $\approx$ 1--2 \AA), our 8- and 24-f.u. simulations are useful for the bond lifetime analysis}.
Within the whole range in density (2.65--7.95~g/cm$^3$) and temperature (5,000--100,000~K) that we have considered,
the Si--O bond lifetime is in general longer at lower temperatures and it gradually drops from approximately 1000 to 20 fs, as temperature increases. 
This trend is similar for all densities, while the value of $\tau$ is typically larger at higher densities (i.e., when the system is more squeezed).
Remarkably, the temperature corresponding to a lifetime of 50~fs increases from 15,000~K at 40 GPa to 38,000 K at 1000 GPa (green solid line-triangle curve in Fig.~\ref{fig:phase}),
which is similar to the bonded-to-atomic transition defined by using the anomaly in heat capacity. 
This indicates that we may choose the lifetime of 50-fs as another criteria to define the bonded-to-atomic transition, as it corresponds to a uniform length in time that Si--O bonds can stably exist.
The two approaches consistently suggest a transition boundary that is lower in temperature and more dependent on pressure than previous findings (37,000 K and weakly dependent on pressure),
which were based on an analysis of the shock temperature to infer approximated values for the specific heat $C_V$ along the fused silica and $\alpha$-quartz Hugoniots~\cite{HicksPRL2006}. 

We note that, if choosing $\tau_\text{Si-O}=25$ (or 100)~fs as the criteria for bond dissociation, the corresponding temperatures are higher (or lower) by 1--2$\times$10$^4$~K than when choosing $\tau_\text{Si-O}=50$~fs, while the sensitivity to pressure remains similar, as shown with dotted (or dashed) line-triangles in in Fig.~\ref{fig:phase}.

\begin{figure}[ht]
\centering
\includegraphics[width=0.8\linewidth]{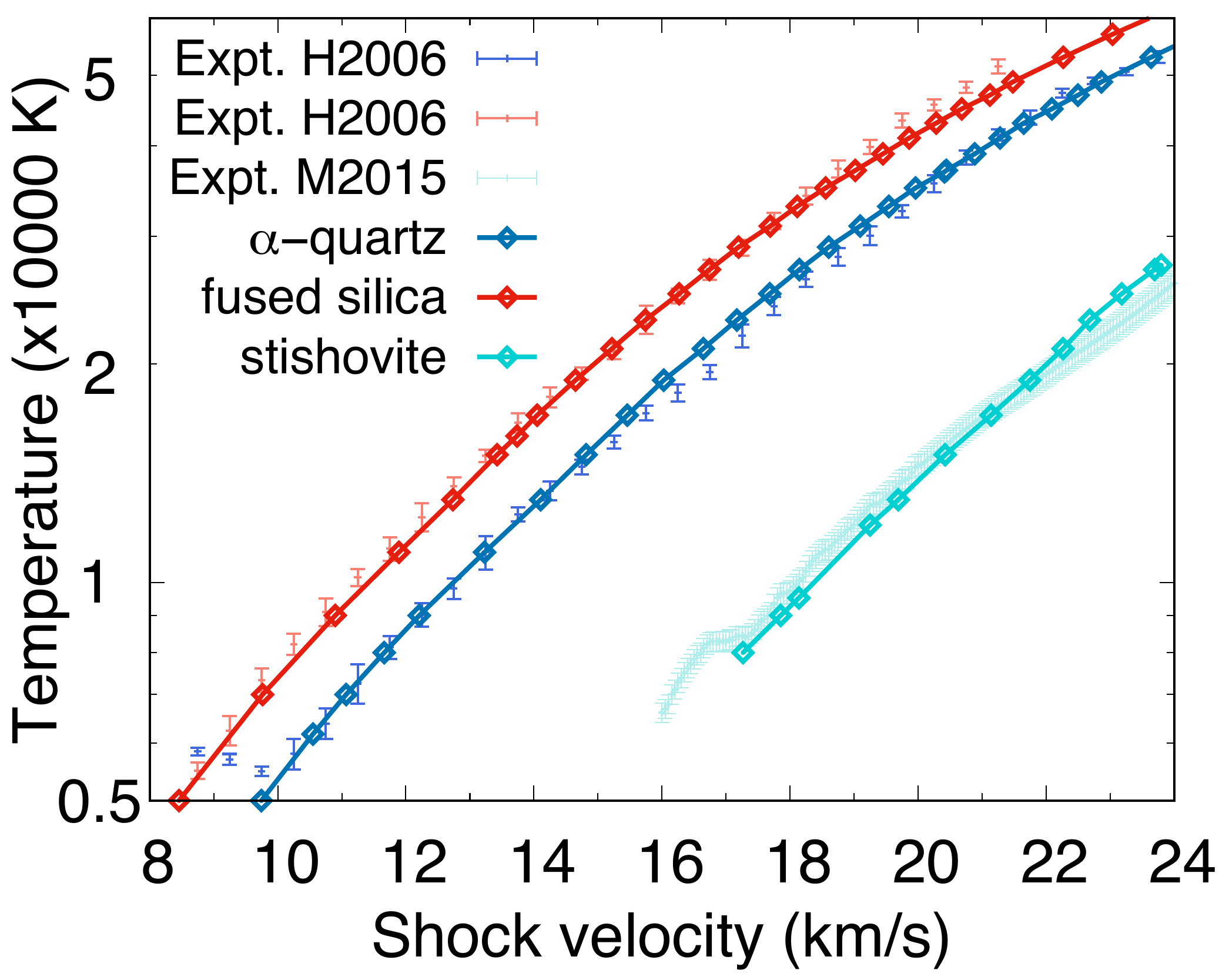}
\caption{Temperature-shock velocity plot of SiO$_2$ Hugoniots in initial forms of fused silica (red), $\alpha$-quartz (blue) and stishovite (turquoise) from our simulations (line-diamonds) compared to experiments
(light-colored symbols)~\cite{HicksPRL2006,MillotSci2015}.}
\label{fig:t-us}
\end{figure}

Figure~\ref{fig:phase} shows overall consistency between our simulations and the measurements in the $T-P$ Hugoniots for $\alpha$-quartz and fused-silica, while
slightly larger differences are noticed for the stishovite near the melting (similar to that found in a previous EOS and DFT-MD study of SiO$_2$ in the fluid regime~\cite{SjostromAIP2017}).
This picture remains similar in the plot of $T$ {\it vs} $u_s$ (Fig.~\ref{fig:t-us}), which are measurable in the experiments and do not reply on the choice of $u_s$--$u_p$ relations (in contrast to $P$ that is calculated by $\rho_0u_su_p$ and thus depends on the $u_s$--$u_p$ relation, which lacks data along the stishovite Hugoniot at 0.2--1.2~TPa.)~\footnote{{\color{black}
We also note differences between computation and experiment along the fused silica Hugoniot at temperatures above 40,000~K, similar to that shown in a $T$--$P$ Hugoniot plot in Ref.~\onlinecite{SjostromAIP2017}. This does not affect the conclusions in this work and can be worthwhile to study in the future.}} 
{\color{black}This implies that the discrepancy in pressure dependence for the bonded-to-atomic transition may be related to the limited thermodynamic space that was probed in the experiment. This poses a higher requirement on the resolution to determine the transition temperature and pressure than that achievable by using the relatively simple models for estimating temperature and isochoric heat capacity.
In contrast, our DFT-MD simulations provide a more complete sampling of temperature conditions along different isochores, which allows straightforward diagnosis of the structure and thermodynamic properties.
However, the accuracy of DFT-MD results is also known to be dependent on the exchange-correlation (XC) functional, pseudopotential, and finite sizes of the simulation cell, which we will address in the following Sec.~\ref{subsec:resultc}. }

\subsection{Validity of the DFT-MD results}\label{subsec:resultc}
\begin{figure}[ht]
\centering
\includegraphics[width=0.9\linewidth]{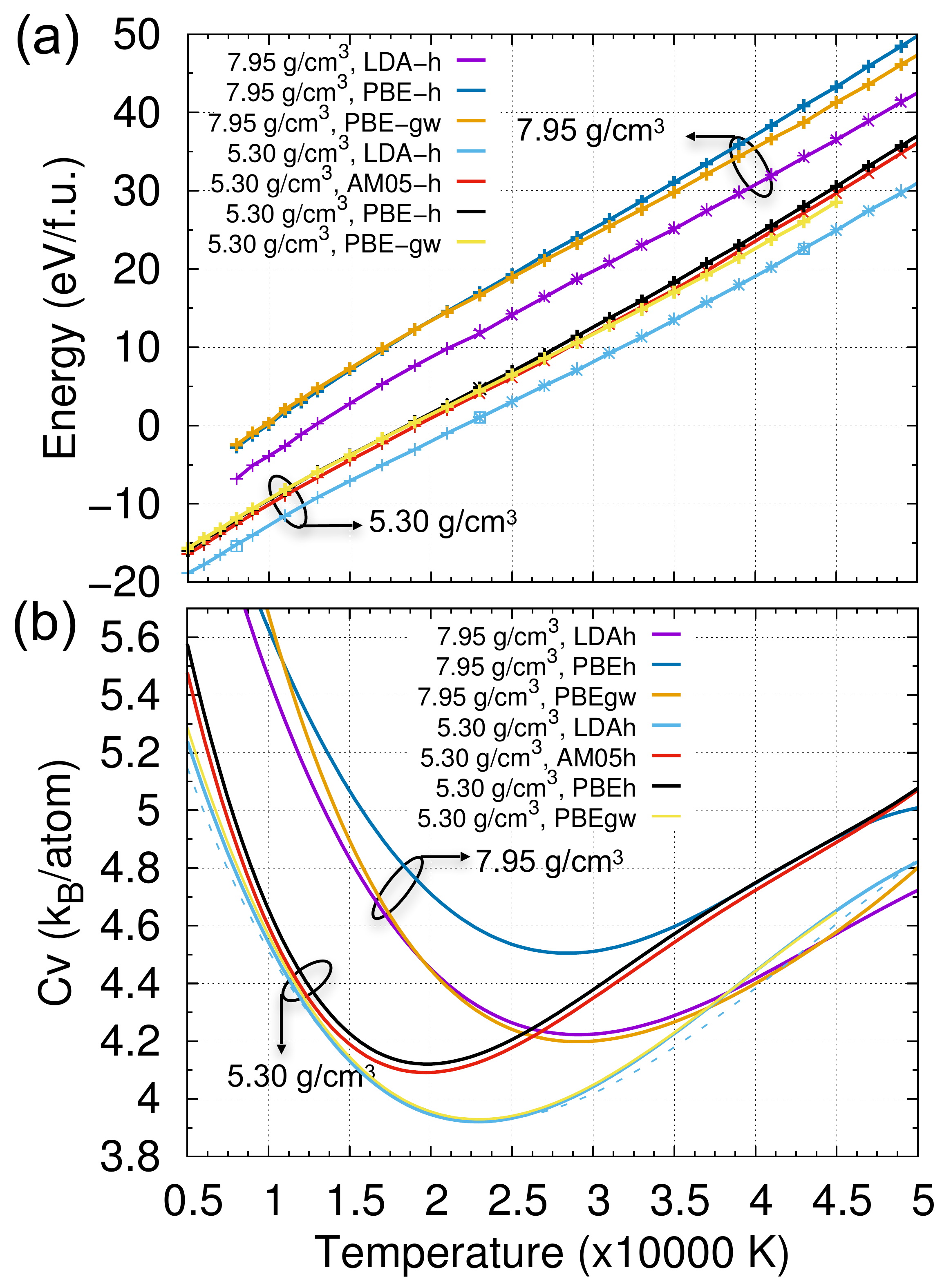}
\caption{Effect of different XC functionals, simulation cell sizes, and pseudopotentials on (a) the internal energy $E(T)$ and (b) the corresponding heat capacity $C_V(T)$ profiles. Notations in (a): ‘+’ denotes 24-f.u. cell simulation; ‘x’ denotes 8-f.u. cell simulation; ‘$\square$’ denotes 4$\times$4$\times$4 k-mesh simulation; ’h’ denotes hard pseudopotentials; ‘gw’ denotes GW-type pseudopotentials. In (b), the blue dashed curve is obtained based on mixed 24-f.u. (at low-T) and 8-f.u. (at high-T) simulations.}
\label{fig:XCpsp}
\end{figure}

We have performed additional calculations by using GGA functionals (AM05 or PBE) and GW-type pseudopotentials to cross-check our findings on the bonded-to-atomic transition based on LDA. The thermodynamic results in $E(T)$ and $C_V(T)$ along two different isochores are shown in Fig.~\ref{fig:XCpsp} and compared to the previous results based on the LDA XC functional and hard PAW method.

Our results show that, within GGA (AM05 or PBE), the transition temperature (as defined by the anomaly in heat capacity) decreases by 3000~K at 5.30~g/cm$^3$ and by 800~K at 7.95~g/cm$^3$, relative to LDA. At the minimum, $C_V$ is higher than LDA because the slope of $E(T)$ curves are larger; this indicates the electron thermal contribution to energies under GGA is larger than that within LDA.
We also find that, when using GW-type pseudopotentials, the slope of $E(T)$ decreases. Therefore, $C_V$ decreases relative to hard-type pseudopotentials, and the transition temperature obtained using PBE is consistent with that obtained using hard pseudopotentials under LDA.
Moreover, we have tried switching from 24-f.u. to 8-f.u. or from $\Gamma$ to $4\times4\times4$ k-mesh for the calculations but observed no meaningful difference in EOS and the transition temperature (see results at 5.30~g/cm$^3$ shown in blue in Fig.~\ref{fig:XCpsp}), which shows our findings are robust and not affected by finite-size effects.

\section{Conclusions}\label{sec:conclusions}
We have performed extensive simulations from first principles and in-depth analysis of the structure, electron density, and thermodynamic properties of liquid silica and provided insights about the nature of the bonded-to-atomic transition in liquid silica.
Our results show smooth internal energy curves as a function of temperature, indicating the transition is likely second-order.
The heat capacity anomaly, which defines the bonded-to-atomic transition, happens at 2--3$\times10^4$~K (1.5--2.5~eV) over the pressure range of 0.1--1~TPa. The transition temperature is lower and more sensitive to pressure than previous estimations~\cite{HicksPRL2006}. These results render a new bonded-to-atomic boundary of liquid silica that overlaps with the conditions of interest to giant-impact simulations~\cite{CanupIracus2004}, which indicates more complex variations (i.e., decrease and then increase with temperatures) in heat capacity than that considered previously~\cite{HicksPRL2006}. This can rebalance the dissipation of irreversible work into temperature and entropy in events of giant impact, necessitating reconsideration of predictions by simulations that are based on empirical EOS models~\cite{Melosh2007,Kraus2012}.

Furthermore, even though the temperature-density grid considered for EOS calculation in this work is relatively sparse, our calculated Hugoniots show overall agreement with experimental results and are similar to previous calculations using alike methods~\cite{ScipioniPNAS2017,SjostromAIP2017}. 
{\color{black}The discrepancies between theory and experiment in the stishovite temperature-pressure Hugoniot near melting, 
together with the previously shown inconsistencies at 1.0--2.5 TPa~\cite{SjostromAIP2017},
also emphasizes the need for further development in both numerical simulations and dynamic compression experiments to 
improve constraints on the phase diagram, EOS, and properties of SiO$_2$ in regions off the Hugoniots of $\alpha$-quartz and fused silica
and elucidate the exotic behaviors affecting matter at extreme condition.
These include simulations that overcome the increased limitations by pseudopotentials and computational cost for reaching convergence at the high density-temperature conditions or go beyond LDA/GGA for the XC functional, as well as more in-depth experimental studies, currently lacking benchmarking $u_s$--$u_p$ data for stishovite between 0.2--1.2 TPa and relying on pyrometry and a grey-body approximation~\cite{QiPoP2015,Falk2014} for temperature estimation.}

\section*{Acknowledgements}
We appreciate Dr. M. Li, Dr. A. Samanta, and Dr. H. Whitley for beneficial discussions and assistance during the research.
This material is based upon work supported by the Department of Energy National Nuclear Security Administration under Award Number DE-NA0003856, the University of Rochester, and the New York State Energy Research and Development Authority.
The Flatiron Institute is a division of the Simons Foundation.
Part of this work was performed under the auspices of the U.S. Department of Energy by Lawrence Livermore National Laboratory under contract number DE-AC52-07NA27344. M.M. acknowledges support from LLNL LDRD project 19-ERD-031. R. J. and E. Z. thank the the Center for Matter at Atomic Pressures (CMAP), a National Science Foundation (NSF) Physics Frontier Center, under Award PHY-2020249. Any opinions, findings, conclusions or recommendations expressed in this material are those of the author(s) and do not necessarily reflect those of the National Science Foundation.

This report was prepared as an account of work sponsored by an agency of the U.S. Government. Neither the U.S. Government nor any agency thereof, nor any of their employees, makes any warranty, express or implied, or assumes any legal liability or responsibility for the accuracy, completeness, or usefulness of any information, apparatus, product, or process disclosed, or represents that its use would not infringe privately owned rights. Reference herein to any specific commercial product, process, or service by trade name, trademark, manufacturer, or otherwise does not necessarily constitute or imply its endorsement, recommendation, or favoring by the U.S. Government or any agency thereof. The views and opinions of authors expressed herein do not necessarily state or reflect those of the U.S. Government or any agency thereof. 



%

\end{document}